\documentclass[10pt,conference]{IEEEtran}

\ifCLASSINFOpdf

\else

\fi
\usepackage[colorinlistoftodos]{todonotes}
\usepackage{amssymb}
\usepackage{amsmath,epsfig,amssymb,verbatim}
\usepackage{cite}
\usepackage[caption=false]{subfig}
\usepackage{array,algorithm,algorithmic}
\usepackage{amsmath,amsfonts,amssymb}
\usepackage{algorithmic}
\usepackage{algorithm}
\usepackage{array}
\usepackage{textcomp}
\usepackage{stfloats}
\usepackage{url}
\usepackage{verbatim}
\usepackage{cite}
\usepackage[normalem]{ulem}
\usepackage{mathtools}
\usepackage{comment}
\def\BibTeX{{\rm B\kern-.05em{\sc i\kern-.025em b}\kern-.08em
    T\kern-.1667em\lower.7ex\hbox{E}\kern-.125emX}}

\usepackage{tablefootnote}
\usepackage{enumerate}

\DeclareMathOperator{\sinc}{sinc}

\DeclareRobustCommand{\erase}{\bgroup\markoverwith{\textcolor{red}{\rule[.5ex]{2pt}{0.5pt}}}\ULon}

\begin{document}

\title{Time-Frequency Localization Characteristics of the Delay-Doppler Plane Orthogonal Pulse
}

\author{\IEEEauthorblockN{Akram Shafie}
\IEEEauthorblockA{
\!\!\!\!\textit{University of New South Wales}~~\\
Sydney, NSW, Australia\\
akram.shafie@unsw.edu.au}
\and
\IEEEauthorblockN{Jinhong Yuan}
\IEEEauthorblockA{
\textit{University of New South Wales}~~\\
Sydney, NSW, Australia\\
j.yuan@unsw.edu.au}
\and
\IEEEauthorblockN{Nan Yang}
\IEEEauthorblockA{
\textit{Australian National University}~~\\Canberra, ACT, Australia\\
nan.yang@anu.edu.au}
\and
\IEEEauthorblockN{Hai Lin}
\IEEEauthorblockA{
\textit{Osaka Metropolitan University}\!\!\!\!\\Sakai, Osaka, Japan\\
hai.lin@ieee.org}}

\maketitle

\begin{abstract}
The orthogonal delay-Doppler (DD) division multiplexing (ODDM) modulation has recently been proposed as a promising solution for ensuring reliable communications in high mobility scenarios. In this work, we investigate the time-frequency (TF) localization characteristics of the DD plane orthogonal pulse (DDOP), which is the prototype pulse of ODDM modulation. The TF localization characteristics examine how concentrated or spread out the energy of a pulse is in the joint TF domain. We first derive the TF localization metric, TF area (TFA), for the DDOP. Based on this result, we provide insights into the energy spread of the DDOP in the joint TF domain. Then, we delve into the potential advantages of the DDOP due to its energy spread, particularly in terms of leveraging both time and frequency diversities, and enabling high-resolution sensing. Furthermore, we determine the TFA for the recently proposed generalized design of the DDOP. Finally, we validate our analysis based on numerical results and show that the energy spread for the generalized design of the DDOP in the joint TF domain exhibits a step-wise increase as the duration of sub-pulses increases.
\end{abstract}


\section{Introduction}

A novel multi-carrier modulation scheme, known as orthogonal delay-Doppler (DD) division multiplexing (ODDM) modulation, has recently been proposed as a promising solution for achieving reliable communications in high mobility scenarios \cite{2022_TWC_JH_ODDM}. The ODDM modulation couples the modulated signal with the stable and sparse DD domain representation of the doubly selective channel. In doing so, it achieves diversity gains while minimizing interference, resulting in improved communication reliability \cite{2021_WCM_JH_OTFS}. Thanks to its enhanced out-of-band emission performance and orthogonality characteristics, ODDM modulation demonstrates superior bit error rates in comparison to the widely studied DD domain modulation scheme, known as orthogonal time-frequency (TF) space (OTFS) modulation \cite{2017_WCNC_OTFS_Haddani,2022_ComLet_Cheng_OTFSErrorPerformance}.

\begin{figure}[t]
\centering
\hspace{-2mm}
\includegraphics[width=1\columnwidth,height=1.7in]{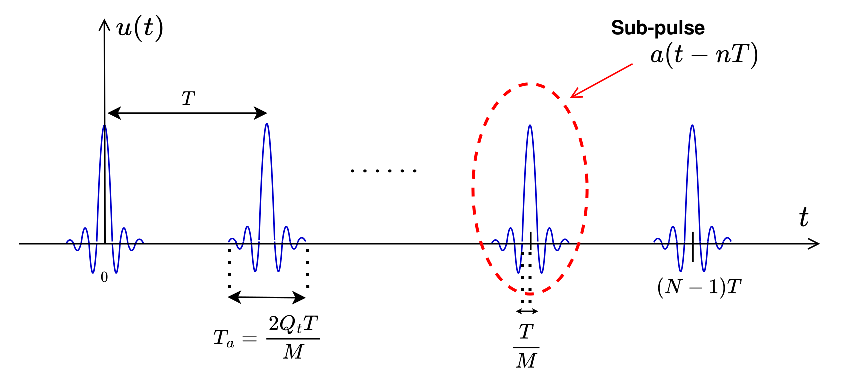}
\caption{Illustration of the DDOP.}\label{Fig:ut}
\end{figure}

At the core of ODDM modulation lies the newly discovered prototype pulse, called the DD plane orthogonal pulse (DDOP). According to \cite{2022_TWC_JH_ODDM}, the DDOP is  generated by concatenating multiple consecutive sub-pulses (see Fig. \ref{Fig:ut}). Due to its potential benefits, \cite{2022_TWC_JH_ODDM,2022_GC_JH_ODDMPulse,2023_ODDM_TCOM} analyzed the orthogonality of the DDOP in various contexts. In particular, \cite{2022_TWC_JH_ODDM} proved the orthogonality of the DDOP with respect to (w.r.t.) the delay and Doppler resolutions of ODDM modulation. Then the proof outlined in \cite{2022_TWC_JH_ODDM} was extended by \cite{2022_GC_JH_ODDMPulse,2023_ODDM_TCOM} to the scenario where
the duration constraint of sub-pulses in the DDOP is relaxed. Despite these endeavors, the literature lacks comprehensive discussions on other properties of the DDOP, particularly its TF localization  characteristics \cite{2017_JSAC_FBMC_forLocalization,2014_ComSurvTut_MCCommunication,1997_Hass_TFLocalization,Book_TheoryofCommunications_Gabor1946}.
This motivates us to delve into the TF localization characteristics of the DDOP, thereby addressing a significant gap in the existing research.

The TF localization characteristics examine how concentrated or spread out the energy of a pulse is in the joint TF domain \cite{2003_TFLocalizationMultiCarrier}. This examination is significant since developing pulses tailored for a specific modulation scheme typically involves managing its TF localization and orthogonality characteristics, while taking into account the properties of the target channel relevant to the modulation scheme \cite{1997_Hass_TFLocalization}.
This development can effectively manage and minimize the inter-symbol-interference (ISI) and/or inter-carrier interference that pulses may encounter from neighboring pulses/signals during the transmission through the target channel.

The TF area (TFA), also referred to as dispersion product, is a well-established metric used in the literature to quantify the TF localization characteristics of a pulse \cite{2017_JSAC_FBMC_forLocalization,2014_ComSurvTut_MCCommunication}. It quantifies the amount of energy spread in the joint TF domain \cite{1997_Hass_TFLocalization}. Due to the constraints imposed by the Heisenberg uncertainty principle, the TFA adheres to a lower limit, known as the Gabor limit \cite{Book_TheoryofCommunications_Gabor1946}. Through deriving the TFA and subsequent analyses, some previous studies investigated the TF localization characteristics of various well-established pulses \cite{2017_JSAC_FBMC_forLocalization,2014_ComSurvTut_MCCommunication,1997_Hass_TFLocalization,Book_TheoryofCommunications_Gabor1946}.
However, none of these studies have explored the TF localization characteristics of the DDOP, which is not surprising since the DDOP has been discovered very recently.

In this work, we investigate the TF localization characteristics of the DDOP. The key contributions of this work are summarized as follows:
\begin{itemize}
\item We derive the TFA for the DDOP while considering the well-known square root-raised-cosine (SRRC) pulse as its sub-pulse;
\item Using this derivation and comparing the energy spread of the DDOP with the pulses used in other established modulation schemes, we provide valuable insights into the energy spread of the DDOP in the joint TF domain;
\item We discuss the potential advantages brought by the energy spread of the DDOP, particularly in terms of leveraging both time and frequency diversities, and enabling high-resolution sensing;
\item We derive the TFA for the recently proposed generalized design of the DDOP in which the duration constraint of sub-pulses is relaxed;
\item We verify our findings using numerical results. We also show that the energy spread for the generalized design of the DDOP in the joint TF domain shows a step-wise rise when 
    the duration of its sub-pulses increases.  
\end{itemize}

\section{DD Plane Orthogonal Pulse and Time Frequency Area}

In this section, we introduce the DDOP and the metric used to quantify the TF localization characteristics of a pulse. 

\subsection{DD Plane Orthogonal Pulse Model}

As shown in Fig. \ref{Fig:ut}, the DDOP is obtained by concatenating $N$ sub-pulses, $a(t)$, which are spaced apart by a duration of $T$. Mathematically, the DDOP is expressed by \cite{2022_TWC_JH_ODDM} 
\begin{align}\label{ut}
u(t)=\sum_{n=0}^{N-1}a(t-nT).
\end{align}
In \eqref{ut}, a sub-pulse $a(t)$ can be the truncated version of any square root-Nyquist (SRN) pulse
that is parameterized by the zero-ISI interval $\frac{T}{M}$, the duration $T_{a}=2Q_{t}\frac{T}{M}$, and the energy $\frac{1}{N}$, where $Q_t$ is a positive  integer \cite{2022_TWC_JH_ODDM}, $N$ is the number of subcarriers in ODDM modulation and $M$ is the number of multi-carrier symbols in the ODDM frame.
We note that satisfying the condition $T_a\ll T$, or equivalently $2Q_t\ll M$, is necessary for the DDOP to maintain strict orthogonality w.r.t. the delay resolution of ODDM modulation \cite{2022_TWC_JH_ODDM,2023_ODDM_TCOM}.

\begin{figure}[t]
\hspace{-5mm}
\includegraphics[width=1.1\columnwidth,height=1.7in]{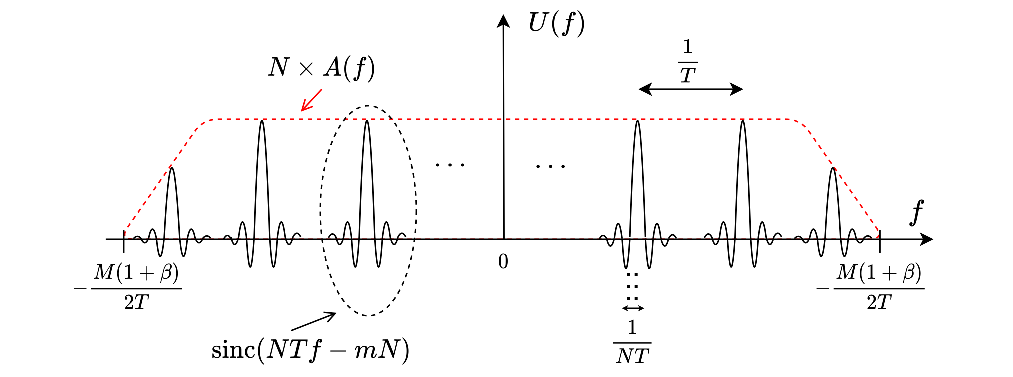}
\caption{Frequency response of the DDOP, with phase terms ignored for the purpose of display.}\label{Fig:Uf} 
\end{figure}

There are several candidates for the SRN pulse in the literature \cite{Book_GaborAnalysis,2001_ComLet_Normal_BTNP}. Without loss of generality, in this work, we consider the well-known SRRC pulse as the SRN pulse \cite{2022_TWC_JH_ODDM}. Under this consideration, the sub-pulse $a(t)$ becomes
\begin{align}\label{at_SRRC}
a(t)=\begin{cases}\sqrt{\frac{M}{N T}} \frac{\sin \left(\pi \frac{Mt}{T}(1-\beta)\right)+ \frac{4 \beta Mt}{T} \cos \left( \frac{\pi M t}{T}(1+\beta)\right)}{ \frac{\pi Mt}{T}\left(1-\left( \frac{4 \beta Mt}{T}\right)^2\right)}, & |t| {\leq}  \frac{T_a}{2}, \\
0, & |t|{>}  \frac{T_a}{2},\end{cases}
\end{align}
where $\beta$ denotes the roll-off factor which can take values ranging from 0 to 1.
As for the frequency response of the DDOP (see Fig.~\ref{Fig:Uf}), it was derived in \cite{2022_GC_JH_ODDMPulse,2023_ODDM_TCOM} as
\begin{align}\label{Uf}
U(f)=&N e^{-j \pi (N-1)Tf} A(f)\notag\\
&\times\sum_{m=-\infty}^{\infty}e^{j \pi (N-1)m}\sinc(NTf-mN),
\end{align}
where 
$A(f)$ is the frequency response of the chosen $a(t)$. For $a(t)$ given in \eqref{at_SRRC}, its $A(f)$ is given by\footnote{We note that the impact of truncating the SRRC pulse in the TD has a marginal impact on its frequency response, particularly when $\beta$ and/or $Q_t$ are relatively large. Considering this and to maintain simplicity, we express the frequency response of the SRRC pulse as the frequency response of the truncated SRRC pulse in \eqref{Af_SRRC}.}
\begin{align}\label{Af_SRRC}
&A(f)=\begin{cases}\sqrt{\frac{T}{MN}}, &\hspace{-3cm}0 \leq|f| \leq \frac{M(1{-}\beta)}{2 T},\\
\sqrt{\frac{T}{2MN}\left(1+\cos\left(\frac{\pi T}{\beta M}\left(|f|{-}\frac{M(1{-}\beta)}{2 T}\right)\right)\right)},\\&\hspace{-3cm}\frac{M(1{-}\beta)}{2 T} \leq|f| \leq \frac{M(1{+}\beta)}{2 T}, \\
0, &\hspace{-3cm}\textrm{otherwise}.\end{cases}
\end{align}

It is interesting to observe from \eqref{ut}, \eqref{Uf}, Fig. \ref{Fig:ut}, and Fig. \ref{Fig:Uf} that, on one hand, the energy of the DDOP is scattered in the time domain (TD) across multiple sub-pulses $a(t)$. On the other hand, its energy is scattered in the frequency domain (FD) across multiple sub-tones $\sinc(NTf)$.

\subsection{Time Frequency Area of a Pulse}

The TFA (or the dispersion product), $\Delta A$, is a classical metric used in the literature to examine the TF localization characteristics of a pulse \cite{2017_JSAC_FBMC_forLocalization,2014_ComSurvTut_MCCommunication}. Particularly, it quantifies the energy spread of a pulse in the joint TF domain. Mathematically, it is defined as the product of the time dispersion, $\Delta T$, and the frequency dispersion, $\Delta F$, given by
\begin{align}\label{Au}
\Delta A=\Delta T\Delta F.
\end{align}

For a pulse $g(t)$, $\Delta T$ and $\Delta F$ are defined as \cite{2014_ComSurvTut_MCCommunication}\footnote{We note that according to the Heisenberg uncertainty principle, time-limited pulses cannot be \textit{strictly} frequency-limited, and frequency-limited pulses cannot be \textit{strictly} time-limited \cite{Book_TheoryofCommunications_Gabor1946,2023_ODDM_TCOM}. Due to this reason, in theory, either $\Delta T$ or $\Delta F$ of a pulse will take the value infinity. Considering this and the fact that practical pulses are inherently time-limited, in this work, we determine $\Delta F$ of a pulse in an essential sense, i.e., $\Delta F$ is determined by ignoring the negligibly very small frequency tails \cite{2023_ODDM_TCOM}.}
\begin{align}\label{DeltaT}
\Delta T=\sqrt{\frac{1}{E_g}\int_{-\infty}^{\infty}(t-\bar{t})^2|g(t)|^2\mathrm{d}t}
\end{align}
and
\begin{align}\label{DeltaF}
\Delta F=\sqrt{\frac{1}{E_g}\int_{-\infty}^{\infty}(f-\bar{f})^2|G(f)|^2 \mathrm{d}f},
\end{align}
respectively, where $E_{g}$ is the energy of $g(t)$, $G(f)$ is the frequency response of $g(t)$, and $\bar{t}$ and $\bar{f}$ represent the mean values of the supports of the pulse in time and frequency, respectively. To be specific, $\bar{t}$ is given by
\begin{align}\label{tbatorg}
\bar{t}=\frac{1}{E_g}\int_{-\infty}^{\infty}t|g(t)|^2 \mathrm{d}t
\end{align}
and $\bar{f}$ is given by
\begin{align}\label{fbarorg}
\bar{f}=\frac{1}{E_g}\int_{-\infty}^{\infty}f|G(f)|^2 \mathrm{d}f.
\end{align}

As a consequence of the Heisenberg uncertainty principle, $\Delta A$ obeys a lower bound known as the \textit{Gabor limit}, $\Delta A\geqslant\frac{1}{4\pi}$, which is attained by the Gaussian pulse~\cite{Book_TheoryofCommunications_Gabor1946,Book_GaborAnalysis}. Typically, a pulse is considered to be well-localized in the joint TF domain (or have the minimum energy spread in the joint TF domain) 
if its $\Delta A$ approaches the Gabor limit \cite{2017_JSAC_FBMC_forLocalization,2014_ComSurvTut_MCCommunication}.

\section{TFA for the DDOP}
\label{Sec:DeltaTForSinc}

In this section, we derive $\Delta A$ for the DDOP, by successively deriving $\Delta T$ and $\Delta F$ for the DDOP. We first start the derivation of $\Delta T$ for the DDOP by expanding \eqref{DeltaT} using \eqref{ut} to obtain
\begin{subequations}\label{DeltaT2}
\begin{align}
\Delta T^2=&\int_{{-}\infty}^{\infty}t^2|u(t)|^2 \mathrm{d}t{-}2\bar{t}\int_{{-}\infty}^{\infty}t|u(t)|^2 \mathrm{d}t\notag\\
&+\bar{t}^{2}\int_{-\infty}^{\infty}|u(t)|^2 \mathrm{d}t\label{DeltaT2a}\\
=&\sum_{n=0}^{N{-}1}\int_{{-}\infty}^{\infty} t^2|a(t{-}nT)|^2 \mathrm{d}t\notag\\
&+\sum_{n=0}^{N{-}1}\sum_{\substack{\dot{n}=0,\dot{n}\neq n}}^{N{-}1}\int_{{-}\infty}^{\infty} t^2|a(t{-}nT)a(t{-}\dot{n}T)|^2 \mathrm{d}t-\bar{t}^{2}.\label{DeltaT2b}
\end{align}
\end{subequations}
We note that the second term in \eqref{DeltaT2b} is zero since $a(t-nT)$ does not overlap with $a(t-\dot{n}T)$ when $n\neq \dot{n}$. Next, we apply
\begin{align}\label{t2shift}
\int_{-\infty}^{\infty} \rho^2|\mathcal{X}(\rho_{\textrm{a}} \rho-\rho_{\textrm{b}})|^2 \mathrm{d}\rho=\int_{-\infty}^{\infty}\rho^2|\mathcal{X}(\rho_{\textrm{a}}\rho)|^2 \mathrm{d}\rho+\frac{\rho^2_{b}}{\rho^2_{\textrm{a}}}E_{\mathcal{X}}
\end{align}
into \eqref{DeltaT2b}, where $\mathcal{X}:\rho\rightarrow \mathcal{X}(\rho)$ is an even function, $\rho\in \mathbb{R}$, $\mathcal{X}(\rho)\in \mathbb{C}$, $E_{\mathcal{X}}=\int_{-\infty}^{\infty}|\mathcal{X}(\dot{\rho})|^2 \mathrm{d}\dot{\rho}$, and $\rho_a,\rho_b\in \mathbb{R}$. By doing so, we further simplify \eqref{DeltaT2b} as
\begin{subequations}\label{T11}
\begin{alignat}{2}
\Delta T^2 =&\sum_{n=0}^{N-1}\left(\int_{-\infty}^{\infty} t^2|a(t)|^2 \mathrm{d}t+ \frac{(n T)^2}{N} \right)-\bar{t}^{2}\label{T11a}\\
=&N\int_{{-}\infty}^{\infty} t^2|a(t)|^2 \mathrm{d}t+\frac{N^2T^2}{12},\label{T11b}
\end{alignat}
\end{subequations}
where \eqref{T11b} is obtained by using the geometric progression formula in \eqref{T11a} and $\bar{t}$ is derived as $\bar{t}=\frac{T}{2}(N-1)$.

Due to the complex nature of the expression for $a(t)$ in \eqref{at_SRRC}, it is extremely difficult, if not impossible, to compute the integral in the first term in \eqref{T11b}. Despite so, on one hand, by carefully observing the first term in \eqref{T11b}, we find that it represents $\Delta T^2$ of the sub-pulse $a(t)$. On the other hand, $\Delta T$ of any pulse is upper bounded by its duration \cite{2014_ComSurvTut_MCCommunication,1997_Hass_TFLocalization}. Considering these, we obtain the upper bound on the first term in \eqref{T11b} as
\begin{align}\label{T111_A}
N\int_{-\infty}^{\infty} t^2|a(t)|^2 \mathrm{d}t &\leqslant T^2_a. 
\end{align}
Next, we substitute \eqref{T111_A} into \eqref{T11b} and further simplify it by considering $T_a\ll T$ and $N$ is sufficiently large. By doing so, we finally obtain 
\begin{align}\label{DeltaT4}
\Delta T
\approx\frac{NT}{\sqrt{12}},
\end{align}
which completes the derivation of $\Delta T$ for the DDOP.

We next derive $\Delta F$ for the DDOP. To begin with, we expanding \eqref{DeltaF} using \eqref{Uf} to obtain
\begin{align}\label{DeltaF2}
\Delta F^2
=&N^2\bigg(\sum_{m=-\infty}^{\infty}\int_{-\infty}^{\infty} f^2|A(f) \sinc(NTf{-}mN)|^2 \mathrm{d}f\notag\\
&\hspace{-7mm}+\sum_{m=-\infty}^{\infty}\sum_{\substack{\dot{m}=-\infty,\dot{m}\neq m}}^{\infty}\int_{-\infty}^{\infty} f^2|A(f)|^2\Big|\sinc(NTf-mN)\notag\\
&\times\sinc(NTf{-}\dot{m}N)\Big|\mathrm{d}f\bigg)-\bar{f}^{2}.
\end{align}
For sufficiently large $N$, the second term in \eqref{DeltaF2} can be approximated as zero, since the energy of $\sinc(NTf)$ concentrated beyond its $\frac{N}{2}$-th zero-crossing point becomes negligible. Also, $\bar{f}$ in \eqref{DeltaF2} can be derived as zero based on the even symmetry of $|U(f)|$. We next note that it is extremely difficult to analytically compute the integral in the first term in \eqref{DeltaF2} for $A(f)$ given in \eqref{Af_SRRC}. Considering this, we first approximate \eqref{DeltaF2} as
\begin{align}\label{DeltaF3}
\Delta F^2\approx
&N^{2}\!\!\!\sum_{m=-\infty}^{\infty}\left|A\left(\frac{m}{T}\right)\right|^2
\int_{-\infty}^{\infty}f^2|\sinc(NTf-mN)|^2 \mathrm{d}f.
\end{align}
We clarify that the impact of the approximation in \eqref{DeltaF3} is marginal due to fact that (i) the frequency range where a considerable amount of energy of $\sinc(NTf)$ is concentrated is relatively small as compared to the span of $A(f)$ and (ii) $A(f)$ varies only within $\frac{M(1{-}\beta)}{2 T} \leq|f| \leq \frac{M(1{+}\beta)}{2 T}$.

\begin{figure*}[t]
\centering
\hspace{-5mm}
\includegraphics[width=1.4\columnwidth]{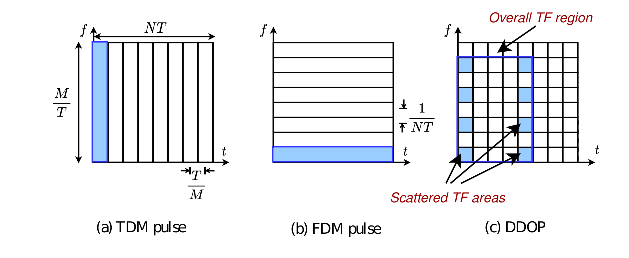} \vspace{-3mm}
\vspace{-3mm}
\caption{Simplified occupancy of the TDM pulse, the FDM pulse, and the DDOP in the TF plane.}\label{Fig:Loc_Compare}
\end{figure*}

Next, applying \eqref{t2shift} into \eqref{DeltaF3}, we further simplify \eqref{DeltaF3} as
\begin{subequations}\label{DeltaF4}
\begin{align}
\Delta F^2 \approx& N^2\sum_{m=-\infty}^{\infty}\left|A\left(\frac{m}{T}\right)\right|^2 \int_{-\infty}^{\infty} f^2|\sinc(NTf)|^2\mathrm{d}t\notag\\
&+N^2\sum_{m=-\infty}^{\infty}\left|A\left(\frac{m}{T}\right)\right|^2 \left(\frac{mN}{NT}\right)^2 \frac{1}{NT}\label{DeltaF4a}\\
\approx& \frac{1}{2\pi^2 T^2}\sum_{m=-\infty}^{\infty}\left|A\left(\frac{m}{T}\right)\right|^2\frac{1}{T}\notag\\
&+N\sum_{m=-\infty}^{\infty}\left(\frac{m}{T}\right)^2
\left|A\left(\frac{m}{T}\right)\right|^2\frac{1}{T}.\label{DeltaF4b}
\end{align}
\end{subequations}
We note that \eqref{DeltaF4b} is obtained by simplifying the first term in \eqref{DeltaF4a} while considering that the energy concentration of $\sinc(NTf)$ beyond its $\frac{N}{2}$-th zero-crossing point is negligible. We then apply the mathematical identify
\begin{align}\label{XXXX}
\lim_{\Delta \rho\rightarrow 0} \sum_{i\in \mathbb{Z}}\mathcal{X}(i\Delta \rho)\Delta \rho=\int_{-\infty}^{\infty}\mathcal{X}(\rho)d\rho
\end{align}
into \eqref{DeltaF4b} to approximate \eqref{DeltaF4b} as
\begin{align} \label{DeltaF45}
\Delta F^2\approx\frac{1}{2\pi^2 T^2}\int_{-\infty}^{\infty}\left|A(f)\right|^2 \mathrm{d}f{+}N\int_{-\infty}^{\infty}f^2\left|A(f)\right|^2 \mathrm{d}f.
\end{align}
The second term in \eqref{DeltaF45} can be derived using $A(f)$ in \eqref{Af_SRRC} as
\begin{align}\label{DeltaFXX}
N\int_{-\infty}^{\infty}f^2\left|A(f)\right|^2 \mathrm{d}f&=\frac{M^2}{12T^2}{+}\frac{(\pi^2{-}8)M^2\beta^2}{4\pi^2T^2}.
\end{align}
Thereafter, we find that the first term in \eqref{DeltaF45} is very small compared to \eqref{DeltaFXX}. Considering this, we obtain $\Delta F$ for the DDOP as
\begin{align}\label{DeltaF5}
\Delta F\approx\frac{M}{T}\sqrt{\frac{1}{12}{+}\frac{(\pi^2{-}8)\beta^2}{4\pi^2}}.
\end{align}
Finally, by substituting \eqref{DeltaT4} and \eqref{DeltaF5} into \eqref{Au}, we obtain $\Delta A$ for the DDOP as
\begin{align}\label{DeltaA_DDOP}
\Delta A\approx\frac{MN}{12}\sqrt{1{+}\frac{3(\pi^2{-}8)\beta^2}{\pi^2}}.
\end{align}

\section{Discussion and Remarks}\label{Sec:Disc}

\subsection{Discussion}\label{Sec:Discussion}

We observe from \eqref{DeltaA_DDOP} that $\Delta A$ for the DDOP is very high compared to the Gabor limit of $\frac{1}{4\pi}$. This indicates that the DDOP is not well-localized in the joint TF domain \cite{1997_Hass_TFLocalization,2017_JSAC_FBMC_forLocalization,Book_TheoryofCommunications_Gabor1946}.

To clearly understand the reasons behind the relatively high $\Delta A$ for the DDOP, we provide a simplified schematic illustration in Fig. \ref{Fig:Loc_Compare}. This figure shows the specific TF regions where the energy of the DDOP, and the pulses used in time division multiplexing (TDM) and frequency division multiplexing (FDM) schemes is concentrated within a TF region bounded by the TD resouce of $NT$ and the FD resource of $\frac{M}{T}$, where $M=4$ and $N=2$ \cite{2023_ODDM_TCOM}.\footnote{For the sake of simplicity, Fig. \ref{Fig:Loc_Compare} only displays the TF regions where the energy of pulses are predominantly concentrated \cite{2023_ODDM_TCOM}.}

For the pulses used in TDM and FDM schemes, their energy is concentrated within a single non-scattered region in the TF plane (see Figs. \ref{Fig:Loc_Compare}(a) and (b)).
As for the dimensions of this non-scattered region, one dimension is very low and the other dimension tends to be very large as these two dimensions have an inverse relationship as per the Heisenberg uncertainty principle \cite{Book_TheoryofCommunications_Gabor1946}.
These lead to the following: In the case of the pulse used in the TDM scheme, $\Delta T$ is relatively small and $\Delta F$ is relatively high, which results in a relatively small $\Delta A$. Differently, in the case of the pulse used in the FDM scheme, $\Delta T$ is relatively high and $\Delta F$ is relatively small, which also results in a relatively small $\Delta A$.

In contrast to TDM and FDM pulses, the energy of the DDOP is scattered in the TD across multiple sub-pulses $a(t)$, as shown in Fig. \ref{Fig:ut}. Also, its energy is scattered in the FD across multiple sub-tones $\sinc(NTf)$, as shown in Fig. \ref{Fig:Uf}. Due to these, the TF region occupied by the DDOP pulse in the TF plane, consists of multiple \textit{small scattered TF areas} that are located far apart (see Fig. \ref{Fig:Loc_Compare}(c)) \cite{2023_ODDM_TCOM}. As a result of this, the energy spread of the DDOP is governed by the dimensions of the \textit{overall TF region} that encompasses all the small scattered TF areas (see the region enclosed by the blue solid line in Fig.~\ref{Fig:Loc_Compare}(c)).
For this overall TF region, both of its dimensions are relatively large, leading to relatively high $\Delta T$ and $\Delta F$, and indicating that the DDOP has wide energy spread across both the TD and the FD. Consequently, $\Delta A$ for the DDOP is relatively high.

\subsection{Remarks}\label{Sec:Remarks}

\subsubsection{Harnessing Time and Frequency Diversities with DDOP}

Despite the relatively high $\Delta A$, $\Delta T$, and $\Delta F$ for the DDOP, it is important to note that wide energy spread across both the TD and the FD is essential for the DDOP \cite{2022_TWC_JH_ODDM}. This is because that the primary objective of the DDOP is to effectively interact with the sparse DD domain representation of doubly selective channels such that it can harness both time diversity and frequency diversity \cite{2021_WCM_JH_OTFS}. Achieving this objective necessitates a pulse with wide energy spread in both the TD and the FD, where a wide energy spread in the TD is essential to harness time diversity and a wide energy spread in the FD is crucial to reap frequency diversity, which align exactly with the characteristics exhibited by the DDOP.

\subsubsection{Sensing with DDOP}

It is important to note that the wide energy spread across both the TD and the FD of the DDOP makes it well suited for high-resolution sensing \cite{2021_WCM_JH_OTFS,2023_ODDM_TCOM,2020_TWC_Lorenzo_OTFSforJSAC}. In particular, the goal of sensing is to accurately estimate \textit{both the range and the velocity of an object} \cite{2021_WCM_JH_OTFS,2023_ODDM_TCOM}. The range estimation relies on the delay of the backscattered pulse, while the velocity estimation relies on the Doppler shift (or phase variation) of the backscattered pulse.
Thus, we note that estimating the range of the object with fine time resolution necessitates the backscattered pulse to exhibit a wide energy spread in the FD, while estimating the velocity of the object with fine frequency resolution requires the backscattered pulse to show a wide energy spread in the TD.

When using pulses such as those used in TDM schemes for sensing, the range of the object can be estimated with fine resolution since pulses used in TDM schemes exhibit a wide energy spread in the FD. Nonetheless, estimating the velocity of the object with fine resolution can pose challenges as pulses used in TDM schemes  lack a wide energy spread in the FD. Conversely, if pulses like those used in FDM schemes are employed for sensing, the Doppler of the object can be estimated with fine resolution since pulses used in FDM schemes exhibit a wide energy spread in the TD. Nonetheless, estimating the delay of the object with fine resolution can be challenging because pulses used in FDM schemes lack a wide energy spread in FD. 

In contrast to pulses used in TDM and FDM schemes, the DDOP enables the simultaneous estimation of both the range and the velocity of an object with fine time and frequency resolutions, due to its wide energy spread in both the TD and the FD. This makes the DDOP highly suitable for high-resolution sensing.


\section{TFA for the Generalized Design of DDOP}\label{Sec:GDDOP}

We note that the DDOP described in \eqref{ut} is formed by concatenating $N$ sub-pulses, and the duration of each sub-pulse is constrained by $T_a\ll T$ to ensure that the orthogonality conditions for the DDOP are met appropriately. However, \cite{2023_ODDM_TCOM} recently proposed a generalized design of the DDOP, referred to as ``general DDOP'', where the duration constraint $T_a\ll T$ of sub-pulses is relaxed, while introducing cyclic prefix (CP) and cyclic suffix (CS) to the pulse. According to \cite{2023_ODDM_TCOM}, the general DDOP is formulated as
\begin{align}\label{ut_general}
\tilde{u}(t)=\sum_{n=-D}^{N-1+D}\tilde{a}\left(t-(n+D)T\right),
\end{align}
where $\tilde{a}(t)$ denotes the sub-pulse without the duration constraint $T_a\ll T$ and $D=\lceil\frac{T_a}{T}\rceil$ with $\lceil\cdot\rceil$ denoting the ceil operation.

Following the steps similar to those presented in Section \ref{Sec:DeltaTForSinc}, we can analyze $\Delta A$ of the general DDOP. Specifically $\Delta A$ of the general DDOP is derived as
\begin{align}\label{DeltaA_GDDOP}
\Delta A\approx\frac{M(N+2D)}{12}\sqrt{1{+}\frac{3(\pi^2{-}8)\beta^2}{\pi^2}}.
\end{align}

\section{Numerical Results}\label{Sec:Numerical}

In this section, we present numerical results to evaluate our analytical expressions. Unless specified otherwise, we consider a normalized $T$, i.e., $T=1$, a roll-off factor $\beta=0.1$, and let $Q_t\approx 0.05\times M$ to ensure that the duration condition $T_a\ll T$ is met for the considered DDOP in \eqref{ut}. For simulation results, we truncate the energy of $\sinc(NTf)$ at its $100$th zero-crossing point.

\begin{figure}[t]
\centering\subfloat[$\Delta A$ versus $\beta$]{ \includegraphics[width=0.9\columnwidth]{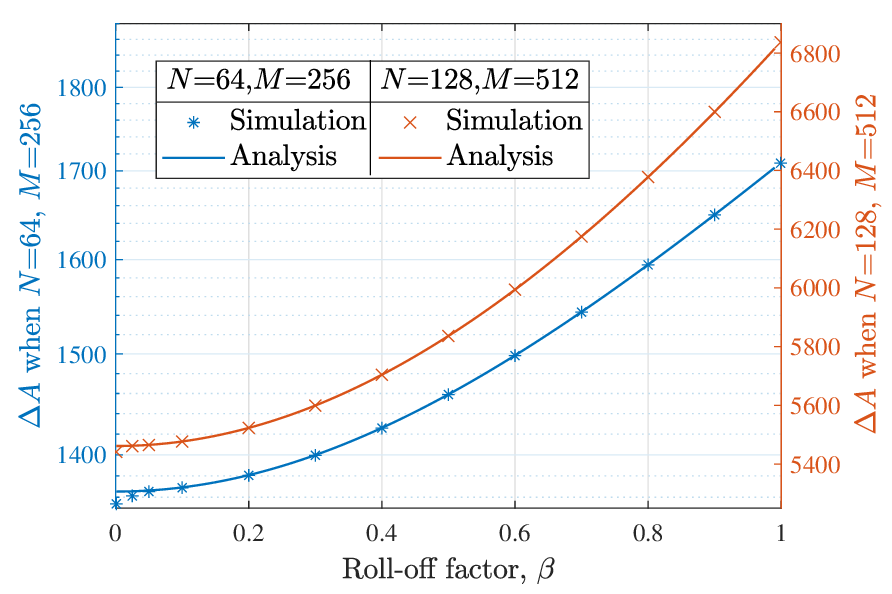}}\hspace{2mm}
\subfloat[$\Delta T$ versus $\beta$]{ \includegraphics[width=0.9\columnwidth]{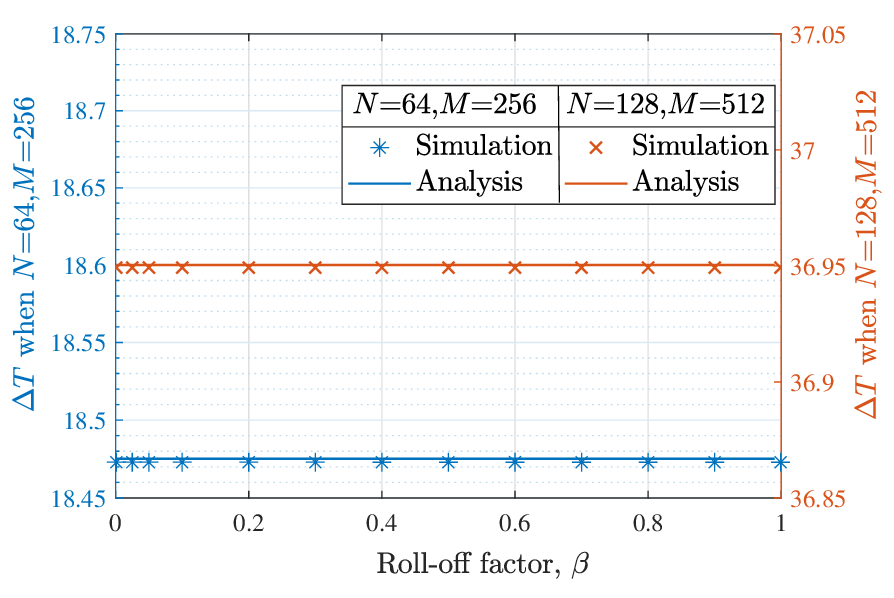}}
\hspace{0 mm}
\subfloat[\label{2b}$\Delta F$ versus $\beta$]{\includegraphics[width=0.9\columnwidth]{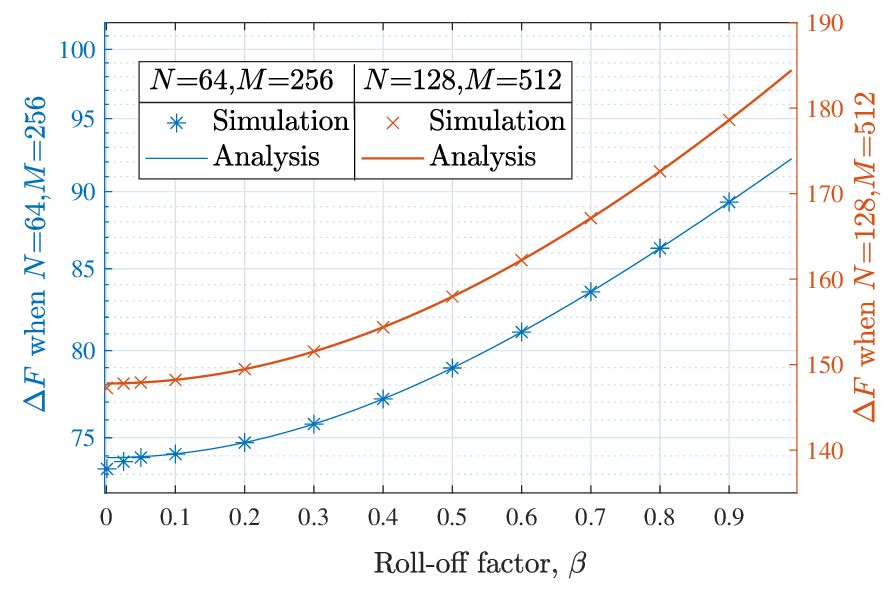}}
\caption{The values of $\Delta A$, $\Delta T$, and $\Delta F$ versus the roll-off factor, $\beta$, for the DDOP considered in \eqref{ut}.}\label{Fig:Deltas_ut_Compare}\vspace{-2mm}
\end{figure}

In Fig. \ref{Fig:Deltas_ut_Compare}, we verify the derived results in Section \ref{Sec:DeltaTForSinc}. To this end, we (i) plot the analytical expressions for $\Delta T$, $\Delta F$, and $\Delta A$, presented in \eqref{DeltaT4}, \eqref{DeltaF5}, and \eqref{DeltaA_DDOP}, respectively, and (ii) numerically calculate $\Delta A$, $\Delta T$, and $\Delta F$ of the simulated DDOP, versus $\beta$ for different values of $M$ and $N$. We observe that for all values of $M$, $N$, and $\beta$, the analytical expressions for $\Delta A$, $\Delta T$, and $\Delta F$ match well with simulation results. For the considered parameters, the maximum percentage difference between the analytical expressions and simulation results for $\Delta A$, $\Delta T$, and $\Delta F$ are $0.8355\%$, $0.0122\%$, and $0.823\%$, respectively. These demonstrate the correctness of our analysis in Section \ref{Sec:DeltaTForSinc}. Here, we clarify that when deriving $\Delta F$, we utilize the frequency response of the SRRC pulse in \eqref{DeltaF3} as the frequency response of the truncated SRRC pulse while ignoring the impact of truncation. This approximation leads to a marginal difference (less than $1\%$) between the analytical expressions and simulation results for $\Delta A$ and $\Delta F$ when $Q_t$ and/or $\beta$ is low. In addition, we note that this marginal difference vanishes as $Q_t$ and/or $\beta$ increases, which further validates our analysis in Section \ref{Sec:DeltaTForSinc}.

\begin{figure}[!t]
\centering
\hspace{-5mm}
\includegraphics[width=0.88\columnwidth]{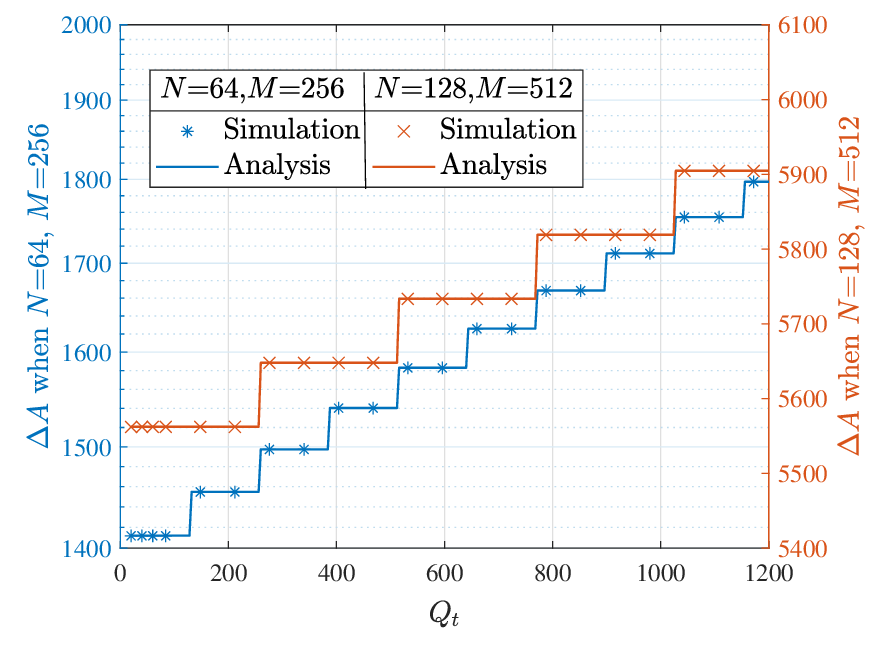}\vspace{-3mm}
\caption{The value of $\Delta A$ versus $Q_t$ for the generalized design of the DDOP given in \eqref{ut_general}.}\label{Fig:Deltas_utgeneral_Compare}\vspace{-3mm}
\end{figure}

In Fig. \ref{Fig:Deltas_utgeneral_Compare},  we verify the derived results for the general DDOP presented in Section \ref{Sec:GDDOP}. To this end, we (i) plot $\Delta A$ given in \eqref{DeltaA_GDDOP} and (ii) numerically calculate $\Delta A$ of the simulated general DDOP, versus $Q_t$ for different values of $M$ and $N$. We observe that for all values of $M$, $N$, and $Q_t$, the analytical expressions for $\Delta A$ match well with simulation results. This proves the correctness of our analysis for the general DDOP in Section \ref{Sec:GDDOP}.
We also observe a step-wise increase in $\Delta A$ in Fig. \ref{Fig:Deltas_utgeneral_Compare} as $Q_t$ increases. This is due to the step-wise increase in the lengths of CP and CS, which occurs in the general DDOP as a result of increasing $Q_t$, since $D=\lceil\frac{T_a}{T}\rceil{=}\lceil\frac{2Q_t}{M}\rceil$.

\section{Conclusion}\label{Sec:Conclusion}

In this work, we studied the TF localization characteristics of the prototype pulse of ODDM modulation, which is called the DDOP. We first derived the TF localization metric, the TFA, for the DDOP. Based on the derived result, we provided useful insights into the energy spread of the DDOP in the joint TF domain. Thereafter, we discussed the potential advantages offered by the DDOP due to its energy spread. Particularly, we highlighted that harnessing both time diversity and frequency diversities necessitates the wide energy spread for the DDOP across both the TD and the FD. We also pointed out that its wide energy spread makes it well-suited for high resolution sensing applications.
We further analyzed the TF localization metric of a recently proposed generalized design of the DDOP. To conclude, we validated our analytical expressions using numerical results.

\vspace{-3mm}

\end{document}